\documentclass[onecolumn,showpacs,superscriptaddress,nobibnotes,nofootinbib,12pt]{revtex4}
\usepackage{amssymb}
\usepackage{amsmath}
\usepackage{epsfig}
\usepackage{graphicx}
\newcommand{\be}{\begin{equation}}
\newcommand{\ee}{\end{equation}}
\newcommand{\bea}{\begin{eqnarray}}
\newcommand{\eea}{\end{eqnarray}}
\newcommand{\nn}{\nonumber}

\newcommand{\g}{\gamma}
\newcommand{\f}{\frac}

\newcommand{\bra}{\langle}
\newcommand{\ket}{\rangle}

\newcommand\lr[1]{{\left({#1}\right)}}

\begin{document}
\title{On multiple scatterings of mesons in hot and cold QCD matter}
\author{Fabio Dominguez}\email{fabio@phys.columbia.edu}
\affiliation{Department of Physics, Columbia University, New York,
NY 10027, USA}
\author{Cyrille Marquet}\email{cyrille@phys.columbia.edu}
\affiliation{Service de Physique Th\'eorique, CEA/Saclay, 91191
Gif-sur-Yvette cedex, France} \affiliation{Department of Physics,
Columbia University, New York, NY 10027, USA}
\author{Bin Wu}\email{bw2246@columbia.edu}
\affiliation{Department of Physics, Peking University, Beijing,
100871, P.R. China} \affiliation{Department of Physics, Columbia
University, New York, NY 10027, USA}
\begin{abstract}

We study the propagation of a color singlet $q\bar q$ pair
undergoing multiple scatterings in hot and cold QCD matter. The
interaction of the dipole with the nucleus or plasma is described
with the McLerran-Venugopalan and Gyulassy-Wang models respectively.
We find identical results when expressed in terms of the saturation
momentum of either the nucleus or the plasma. We compare two kinds
of multiple scatterings, elastic and inelastic with respect to the
target. When allowing the target to scatter inelastically, the
difference with the elastic case is suppressed by a $1/N^2_c$ factor.
We also discuss some implications of our results in the following
situations: the survival probability of quarkonia in a hot medium, the production
of high-$p_T$ heavy mesons in nucleus-nucleus collisions, and the production
of vector mesons in deep inelastic scattering off nuclei.

\end{abstract}


\maketitle
\clearpage
\section{Introduction}

The problem of meson dissociation in QCD matter is of great
importance in relativistic heavy-ion collisions. Indeed one of the
goals of such collisions is to create and understand the quark gluon
plasma (QGP) \cite{rhicexp}, and for many of the possible QGP signatures, the
dissociation of mesons in hot or cold matter is involved at some
stage. Understanding the mechanisms of meson dissociation is crucial
in order to establish a quantitative description of the QGP. It is
therefore important to establish how a fast moving bound state is
broken apart due to the presence of QCD matter.

One of the important experimental signature of the QGP formation in
heavy ion collisions is the suppressed production of high-$p_\perp$
hadrons. In standard medium-induced energy loss calculations
\cite{eloss}, it is assumed that the partons from the hard scattering
hadronize outside of the medium, and therefore that only the energy
loss of partons is responsible for jet quenching. However for heavy
$D$ and $B$ mesons, whose formation time is less than that of pions,
hadronization can happen in the medium, and the problem of meson
dissociation in QCD matter becomes relevant. Perhaps this could help
improve the description of heavy-meson suppression \cite{Adil:2006ra}, which is
underestimated in the standard picture \cite{hqeloss}, even after the
inclusion of collisional energy loss \cite{collEloss}.

Another measurement which is not yet fully understood is the suppressed production
of quarkonia \cite{jpsisup}. Strong $J/\Psi$ suppression has been observed in nucleus-nucleus
collisions by a number of experiments, and various theoretical explanations have been proposed.
The natural explanation invoked is the Debye screening \cite{MatsuiSatz}: a bound state dissociates
because the attractive force between its quark and antiquark is weakened by the
color screening caused by the QGP constituents, thermalized quarks and gluons.

However,
$J/\Psi$ suppression is also seen in hadron-nucleus collisions and it turns out that cold nuclear matter
effects are crucial \cite{klns} (note that the mechanisms of heavy quarkonium production in the vacuum
are already quite involved \cite{JPL}). The suppression seen in nucleus-nucleus collision is likely due to the interplay
between Debye screening, cold nuclear matter effects \cite{QiuVary,QVbetter}, and other phenomena
such as recombination \cite{recomb} or another mechanism for meson dissociation: multiple scatterings
in the QGP. Interestingly enough, it has recently been shown that for infinite-extend matter, multiple scatterings
(collisional dissociation) are a more efficient way of dissociating bound states than Debye screening \cite{bin&fabio}.

Concerning cold nuclear matter effects, it is important to understand them in order
to distinguish to what extend experimental observations in heavy-ion collisions are due to
initial-state or final-state effects. In the case of quarkonium production, cold matter effects
have their own interest, they are relevant for vector meson production in deep inelastic
scattering off nuclei at high energies, where large gluon densities in the nucleus are
probed \cite{vmdis}. This is an important part of the physics program at a future electron-ion
collider \cite{eic}.

In this work we focus on the multiple scatterings. This is the natural mechanism for meson
dissociation in cold nuclear matter, and the dominant one in hot matter for a long enough medium.
Having in mind high-energy mesons, we will work within the eikonal approximation. This is suited to discuss
forward meson production in hadronic collisions involving only cold matter, and high$-p_T$ production in heavy-ion
collisions creating hot matter. We describe the interaction of the mesons with the target nucleus or plasma with the
McLerran-Venugopalan (MV) \cite{mv} and Gyulassy-Wang (GW) \cite{gw} models respectively.
We find that multiple scatterings of mesons are very similar in cold and hot matter, they are controlled
by the saturation momentum, of either the nucleus or the plasma. This scale determines whether the
meson sees a dense or dilute gluon density. We investigate in details the differences between elastic
and inelastic scatterings of the target.

The case where the target scatters elastically has been well studied, it only involves the calculation of a two-point function.
The situation where one allows the target to break up has not received much attention, and this mechanism is the focus of this
paper. It involves the derivation of a four-point function, which is the main technical result of this paper. While several four-point
functions have been obtained in the literature \cite{4ptfunc-jpfr,4ptfunc-jy,4ptfunc-c}, the one computed in this paper is new.
A general algebraic derivation is explained, and in the large $N_c$-limit a diagrammatic derivation is also given.

The plan of the paper is as follows. In Section II, we show how the survival probability of a meson in a plasma and the production of
vector mesons in deep inelastic scattering are related to a two-dipole correlator, in the eikonal approximation. In Section III,
we compute this correlator in the MV model for cold nuclear matter and in Section IV this is done in the GW model for hot QCD matter
with an emphasis on the large $N_c$ limit. Section V is devoted to discussions on possible consequences for the dissociation of quarkonia in the QGP, the suppressed production of heavy-mesons in nucleus-nucleus collisions, and for the production of vector
mesons in electron-ion collisions. Section VI concludes.

\section{Dipole scattering in the eikonal approximation}

In this section we derive the scattering matrix element between two
$q\bar q$ color singlet dipole states. This is relevant for the
calculation of the survival probability of a meson in a plasma, or
the production of vector mesons in deep inelastic scattering. In
both cases, these processes involve the two wave functions
describing the fluctuation of the initial and final states into
$q\bar q$ dipoles, and the $\hat{S}$ matrix element between the
dipole states. We review the formalism in the eikonal approximation.

We shall use light-cone coordinates with the incoming particle being
a right mover. Using light-cone perturbation theory (for an introduction see
\cite{lcpt-bl}) and neglecting higher Fock
components, we write the $q\bar q$ dipole wave function of a meson
or virtual photon with tri-momentum $P=(P^+,P_\perp)$ and
polarization $\lambda:$
\begin{equation}
\left|P,\lambda\right>=\int\frac{d^3p}{(2\pi)^3}\phi^{\lambda}_{h\bar{h}}(p_\perp,z)
\f{\delta_{c\bar{c}}}{\sqrt{N_c}}
\left|p_\perp,z,h,c;P_\perp\!-\!p_\perp,1-z,\bar{h},\bar{c}\right>,\label{WaveFunction}
\end{equation}
where $p=(zP^+,p_\perp)$ denotes the momentum of the quark, $z$ is
the fraction of longitudinal momentum $P^+$ carried by the quark,
$c$ and $\bar{c}$ are color indices, and $h$ and $\bar{h}$ are
polarization indices.

The normalization of the $q\bar q$ state is
\begin{equation}
\left<P^{\prime},\lambda^\prime|P,\lambda\right>=(2\pi)^3
2P^+\delta^{(3)}(P^\prime-P)\delta^{\lambda\lambda^\prime}
\sum\limits_{h\bar{h}}\int\frac{d^2p_\perp dz}{16\pi^3}|\phi^{\lambda}_{h\bar{h}}(p_\perp,z)|^2\ ,
\end{equation}
where the photon wave function is calculated in QED and is not
normalized to unity, while the wave function for the meson is
calculated in various models and required to be normalized to unity, meaning:
\be \sum\limits_{h\bar{h}}\int\frac{d^2p_\perp dz}{16\pi^3}
|\phi^{\lambda}_{h\bar{h}} (p_\perp,z)|^2=1.\ee

The $\hat{S}-$matrix element $\left<P'\right|\hat{S}\left|P\right>$
is
\begin{equation}
S_{fi}=\frac{1}{N_c}\int\f{d^3p}{(2\pi)^3}\f{d^3p\prime}{(2\pi)^3}
\phi_{f}^{*}(p_\perp^\prime,z^\prime)\phi_{i}(p_\perp,z)
S_{dc}^q(p\rightarrow p^\prime)S_{cd}^{\bar q}(P\!-\!p\rightarrow P^\prime\!-\!p^\prime)\ ,
\label{Sfi}
\end{equation}
where $S^{q}$ and $S^{\bar q}$ correspond to the scattering of the
quark and antiquark respectively, and $d$ and $c$ are color indices.
In formula \eqref{Sfi}, the polarization indices of the wave
function are kept implicit, spins are conserved during the eikonal
interaction and are not relevant. In fact the only quantum number
changing is the color. The transverse momenta of the partons also
changes however in the high-energy limit, when the partons propagate
through the hot or cold matter, they have frozen transverse
coordinates and the matrix element depends only of those transverse
coordinates. Therefore it is convenient to introduce the wave
function and scattering matrix element in a mixed representation
$(x_\perp,z)$ with $x_{\perp}\equiv x_{q\perp} - x_{\bar{q}\perp}$:
\be \phi(p_\perp,z)=\int d^2x_\perp \exp
\left\{-ix_\perp\cdot \left(p_\perp-\frac{m_q}{m_q+m_{\bar{q}}}P_{\perp}\right)\right\}
\varphi(z,x_\perp)\ ,\label{mix} \ee
\be S_{dc}^q(p\rightarrow p^\prime)=2\pi\delta(zP^+-z'P^+)\int d^2x_{q\perp}\
e^{i(p_\perp-p'_\perp).x_{q\perp}} W_{dc}(x_{q\perp})\ ,\label{StoW}
\ee where $\left(p_\perp-\frac{m_q}{m_q+m_{\bar{q}}}P_{\perp}\right)$ is the
transverse momentum conjugate to $x_\perp$. The Fourier
transformation \eqref{mix} is defined such that the $P^+$ dependence
in $\varphi$ only enters through $z=p^+/P^+,$ and such that there is
no residual $P_\perp$ dependence in $\varphi(z,x_\perp).$ In formula
\eqref{StoW}, the scattering of the quark is described by the
fundamental Wilson line (see for instance \cite{kw})
\be W[{\cal A}](x_\perp)={\cal P}\exp\lr{ig_S\int dz^+T^c{\cal A}^-_c(z^+,x_\perp)} \ee
where ${\cal P}$ denotes an ordering in $z^+.$ This matrix in color space is a
function of the classical color field ${\cal A}^-$ (we work in the
gauge ${\cal A}^+=0$) describing the QCD matter. The properties of
this color field will be discussed in more detail in the following
section. The scattering of the antiquark is described by the Wilson
line $W^\dagger_{cd}(x_{\bar{q}\perp}).$

This allows us to write \bea S_{fi}&=&2\pi\
2P^+\delta(P^{\prime+}-P^+)M\\M&=&\int \frac{dz}{4\pi} d^2x_{\perp}
d^2X_{\perp} e^{i(P_{\perp}-P^\prime_{\perp})\cdot X_{\perp}}
\varphi_{f}^{*}(z,x_\perp)\varphi_{i}(z,x_\perp)S_{q\bar
q}(x_{q\perp},x_{\bar{q}\perp})\ . \eea The coordinate variables
$x_\perp$ and $X_\perp$ are the dipole size and center of mass
respectively, they are defined in terms of the quark and antiquark
coordinates $x_{q\perp}$ and $x_{\bar{q}\perp}$ in the following
way: \be x_{\perp}=x_{q\perp}-x_{\bar{q}\perp}\ ,\hspace{0.5cm}
X_\perp=\f{m_q x_{q\perp}+m_{\bar q} x_{\bar{q}\perp}}{m_q+m_{\bar
q}}\ . \ee The dipole scattering matrix $S_{q\bar q}$ is a trace of
Wilson lines: \be S_{q\bar q}(x_{q\perp},x_{\bar{q}\perp})=\f1{N_c}
\mbox{Tr}\left[W\left(x_{q\perp}\right)W^{\dagger}\left(x_{\bar{q}\perp}\right)\right]\ .
\ee
Introducing the transverse and longitudinal overlap functions
between the $q\bar q$ color singlet states (for completeness we have
reintroduced the spin indices)
\be
\Phi^T_{fi}(x_\perp)=\f{1}{2}\sum\limits_{\lambda=\pm1}
\int\frac{dz}{4\pi} \sum_{h\bar{h}}
\varphi^{\lambda*}_{h\bar{h},f}(z,x_\perp)
\varphi^{\lambda}_{h\bar{h},i}(z,x_\perp)\ ,\ee
and
\be
\Phi^L_{fi}(x_\perp)=\int \frac{dz}{4\pi} \sum_{h\bar{h}}
\varphi^{0*}_{h\bar{h},f}(z,x_\perp)
\varphi^{0}_{h\bar{h},i}(z,x_\perp)\ ,
\ee
the survival probability of the meson is given by
\begin{equation}
P_{T,L}=\f1{A}\int\f{d^2P_\perp^\prime}{(2\pi)^2}\left<|M|^2\right>\\
=\int d^2x_{\perp} d^2x_{\perp}^\prime
\Phi^{T,L}_{fi}(x_\perp)\Phi^{T,L *}_{fi}(x^\prime_\perp)
\left<S_{q\bar q}(x_{q\perp},x_{\bar{q}\perp}) S_{q\bar
q}(x^{\prime}_{\bar{q}\perp},x^{\prime}_{q\perp})
\right>,\label{Probability}
\end{equation}
where $A=\int d^2X_\perp$ is the cross-sectional area and
$\left<\cdot\cdot\cdot\right>$ represents the medium average which
we shall discuss in the following sections. $x_{q\perp}^\prime$ and
$x_{\bar{q}\perp}^\prime$ are the quark and antiquark transverse
coordinates in the conjugate amplitude:
\be
x^\prime_{q\perp}=X_\perp+\f{m_{\bar q}}{m_q+m_{\bar q}}x^\prime_{\perp}\ ,\hspace{0.5cm}
x^\prime_{\bar{q}\perp}=X_\perp-\f{m_q}{m_q+m_{\bar q}}x^\prime_{\perp}\ .
\ee
We have assumed that
$\left\bra\mbox{Tr}\left[W^F\left(x_{a\perp}\right)W^{F\dagger}\left(x_{b\perp}\right)\right]
\mbox{Tr}\left[W^F\left(x^{\prime}_{b\perp}\right)W^{F\dagger}\left(x^{\prime}_{a\perp}\right)\right]\right\ket$
is independent of $X_\perp$ which we justify in the following calculations.

The diffractive production of vector mesons in deep inelastic
scattering also involves this dipole-dipole correlator, the cross section reads
\begin{equation}
\sigma_{T,L}=\int d^2x_{\perp} d^2x_{\perp}^\prime
\Phi^{T,L}_{V\g}(x_\perp)\Phi^{{T,L}
*}_{V\g}(x^\prime_\perp) \int d^2X_\perp\left<[1-S_{q\bar
q}(x_{q\perp},x_{\bar{q}\perp})][1- S_{q\bar
q}(x^{\prime}_{\bar{q}\perp},x^{\prime}_{q\perp})] \right>\
,\label{cross-section}
\end{equation}
with the appropriate overlap functions $\Phi^\lambda_{V\g}$ between the photon and vector meson wave functions. Note that if we replace the correlator
$\left<S_{q\bar q}(x_{q\perp},x_{\bar{q}\perp})
S_{q\bar q}(x^{\prime}_{\bar{q}\perp},x^{\prime}_{q\perp})\right>,$ by
$\left<S_{q\bar q}(x_{q\perp},x_{\bar{q}\perp})\right>
\left<S_{q\bar q}(x^{\prime}_{\bar{q}\perp},x^{\prime}_{q\perp})\right>,$ then we are only including scattering processes in which the target scatters elastically, this corresponds to exclusive production. However formula \eqref{cross-section} includes the possibility of
the target dissociating.

\begin{figure}
\begin{center}
\includegraphics[width=14cm]{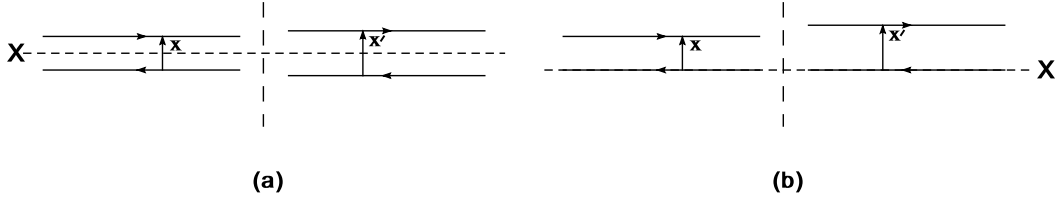}
\end{center}
\caption{Two relevant cases which fix the center of mass of the dipoles $X_\perp:$
quarkonia (a) and heavy mesons (b). The dipoles have the same center of mass in the amplitude and the conjugate amplitude because we are integrating over the meson transverse momentum in the final state \eqref{Probability}.} \label{TwoCaseofMC}
\end{figure}

The purpose of this paper is to calculate
$\left<S_{q\bar q}(x_{q\perp},x_{\bar{q}\perp})
S_{q\bar q}(x^{\prime}_{\bar{q}\perp},x^{\prime}_{q\perp})\right>,$ and to compare the result with the product of two-point functions $\left<S_{q\bar q}(x_{q\perp},x_{\bar{q}\perp})\right>
\left<S_{q\bar q}(x^{\prime}_{\bar{q}\perp},x^{\prime}_{q\perp})\right>.$ We shall
compute the medium average in the MV model for cold matter, and in the GW model for hot matter, the calculation is essentially the same. The derivation of this 4-point function is the main result of this paper. As pictured in Fig.~\ref{TwoCaseofMC}, we will be interested in two particular cases:
\begin{itemize}
\item (a) the quarkonium or vector meson case $m_q=m_{\bar q}$ meaning:
\be X_\perp=\f12\left(x_{q\perp}+x_{\bar{q}\perp}\right)=\f12
\left(x^\prime_{q\perp}+x^\prime_{\bar{q}\perp}\right)\ ,\ee
\item (b) the heavy meson case $m_{\bar q}\gg m_q,$ this puts two Wilson lines at the same coordinate, which greatly simplifies the calculation:
\be X_\perp=x_{\bar{q}\perp}=x^\prime_{\bar{q}\perp}\ .\ee
\end{itemize}
Note that we shall only consider quantities integrated over $P^\prime_\perp,$ the transverse momentum of the meson in the final state. This is why the two center of masses in the amplitude and the conjugate amplitude are the same (see formula \eqref{Probability}). However, the
$P^\prime_\perp$ dependence could also be obtained from the results derived in the following.

\section{Multiple scattering of a color singlet dipole in the McLerran-Venugopalan model}

In the Color Glass Condensate framework, the low energy partons of a
nuclear wave function, those relevant in high-energy processes, are
described by classical color fields. The MV model \cite{mv} is a model for the
distribution of color charges which generate the field. It is a
Gaussian distribution whose variance is the transverse color charge
density squared along the projectile's path $\mu^2(z^+).$ The only parameter
is the saturation momentum $Q_s,$ with $Q_s^2$ proportional to the integrated
color density squared.

\subsection{Introduction to the MV model}

The nuclear average of a function of Wilson lines $f[{\cal A}]$
reads \be \bra f[{\cal A}]\ket=\int D\rho\ \exp\lr{-\int d^2xd^2y
dz^+ \f{\rho_c(z^+,x)\rho_c(z^+,y)}{2\mu^2(z^+)}} f[{\cal A}]\ ,
\label{MV}\ee where the color charge $\rho_c$ and the field ${\cal
A}_c^-$ obey the Yang-Mills equation
\be -\nabla^2{\cal A}_c^-(z^+,x)=g_S\rho_c(z^+,x)\ .\label{YM} \ee
The MV distribution is a Gaussian distribution, therefore one can
compute any average by expanding the Wilson lines in powers of
$g_S\rho_c$ and using Wick's theorem. All correlators of $\rho$'s
can be written in terms of \be \bra\rho_c(z^+,x)\rho_d(z'^+,y)\ket=
\delta_{cd}\delta(z^+-z'^+)\delta^{(2)}(x-y)\mu^2(z^+)\ . \ee Note
that we dropped the $\perp$ indices denoting transverse vectors, in
order to get lighter expressions in the following.

Inverting equation \eqref{YM} gives ${\cal A}_c^-$ in terms of
$\rho_c:$ \be {\cal A}_c^-(z^+,x)=g_S\int d^2z\ G(x-z)\rho_c(z^+,z)\
,\hspace{0.5cm}G(x)=\!\int\limits_{|k|>\Lambda_{QCD}}\!\f{d^2k}{{(2\pi)^2}}
\f{e^{ik\cdot x}}{k^2}\ ,\ee where $G$ is the two-dimensional
massless propagator. After expanding the Wilson lines and applying Wick's
theorem, every contribution is a product of correlators like
\bea g_S^2\bra{\cal A}_c^-(x^+,x){\cal
A}_d^-(y^+,y)\ket &=&\delta_{cd}\delta(x^+-y^+)\mu^2(x^+)g_S^4\int
d^2z\ G(x-z)G(y-z)\\&\equiv&\delta_{cd}\delta(x^+-y^+)\mu^2(x^+)L_{xy}\label{EnseAver}
\eea times a trace of color matrices. The color algebra is the
difficult part to deal with.

\subsection{The dipole-dipole correlator}

\begin{figure}[t]
\begin{center}
\includegraphics[width=14cm]{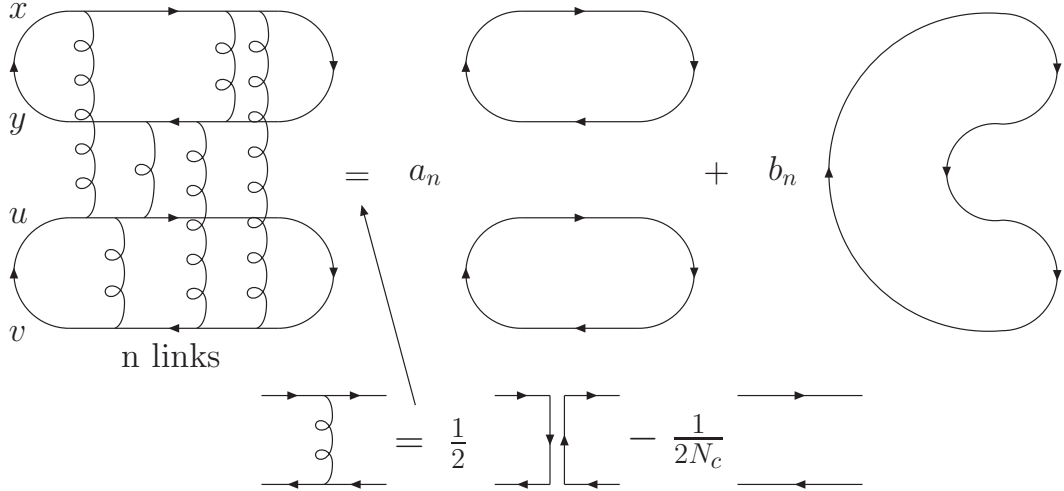} \caption{On the left is a representation of the
correlator \eqref{dipdipcor} with an horizontal line for each Wilson line. Each vertical
link corresponds to a $\bra{\cal A A}\ket$ correlator \eqref{EnseAver} in the
$g_S{\cal A}$ expansion. The figure shows how, when using the Fierz identify \eqref{fierz}, the color structure can be obtained in terms of the two coefficients $a_n$ and $b_n.$ Including
diagrams in which a Wilson line is connected to itself, one gets formula \eqref{ssanbn}.}
\label{Fig:nlink}
\end{center}
\end{figure}

We now compute the following average
\be
\left<S_{q\bar q}(x,y)S_{q\bar q}(u,v)\right>\ .\label{dipdipcor}
\ee
Let's represent each $W$ by a
line along the $z^+$ direction, at a given transverse coordinate. Due
to their respective $z^+$ ordering, $W$'s are oriented to the right
and $W^\dagger$'s to the left. Due to the color structure, the lines are
connected as shown in the left diagram of Fig.~\ref{Fig:nlink}. Let's expand
the Wilson lines in and use Wick's theorem. Graphically, every
$\bra{\cal A A}\ket$ correlator (given by \eqref{EnseAver}) can be
represented by a gluon link between two Wilson lines, at the
relevant time $z^+,$ and the factor associated with it is
$\mu^{2}(z^+)L_{xy}$ with $x$ and $y$ the transverse positions of the
Wilson lines. This comes with a minus sign if the line is between
two (anti)quarks.

A first class of diagrams, easy to deal with (and shown later in
Fig.~\ref{FourTypeScat}(a)), are those where a Wilson line is
connected to itself. Then the contribution is (for instance)
$-C_F\mu^{2}L_{xx}/2,$ it is color singlet and factorizes. The 1/2
is due to the $z^+$ ordering in a single Wilson line. Summing
contributions with an arbitrary number $n$ of such links (and diving
by $n!$ in order not to overcount diagrams) yields \be
T=e^{-\f{C_F}2\mu^2(L_{xx}+L_{yy}+L_{uu}+L_{vv})}\ .\ee
The second class of diagrams are those where two different Wilson
lines are connected, as shown in the figure. Let's simplify their
color structure by using \be
T^a_{ij}T^a_{kl}=\f12\delta_{il}\delta_{jk}-\f1{2N_c}\delta_{ij}\delta_{kl}
\label{fierz}\ee which is represented by the lower diagram in
Fig.~\ref{Fig:nlink}. In doing so, only two topologies can be
obtained, the $(x,y)$ and $(u,v)$ color connections can be flipped
into $(x,v)$ and $(u,y),$ or not, and any diagram is the sum of two
contributions.

Therefore one can write \be \bra S_{q\bar
q}(x,y)S_{q\bar q}(u,v)\ket=
\f{T}{N_c^2}\sum_{n=0}^\infty\int\limits_{z_1^+<\dots<z_n^+}\!\left[
N_c^2a_n(z_1^+,\dots,z_n^+)+N_cb_n(z_1^+,\dots,z_n^+)\right]\ .
\label{ssanbn}\ee
The diagrams have been classified by the number $n$ of links they
contain. At each order, the sum of diagrams is made of two
contributions. The ones where the color structure is not changed
(whose sum is denoted $a_n$) are multipled by $N_c^2$ and the ones
where the color structure is flipped (whose sum is denoted $b_n$)
are multipled by $N_c.$ This is illustrated in Fig.~\ref{Fig:nlink}.
The values of $a_n$ and $b_n$ can be obtained by iteration. The
derivation is given in the appendices A of \cite{4ptfunc-jpfr} and \cite{HiroFujii}. The
sum of diagrams at order $n$ is given by the sum of diagram at order
$n-1,$ times the factor for the $n$-th link. There are six
possibilities to add the $n$-th link on each of the two classes of
diagrams. This gives \be
\lr{\begin{array}{cc}a_n\\b_n\end{array}}=\mu^{2}(z_n^+)M
\lr{\begin{array}{cc}a_{n-1}\\b_{n-1}\end{array}} \ee where the
$2\times2$ matrix is given by \be
M=\lr{\begin{array}{cc}(L_{xy}+L_{uv})C_F+\f1{2N_c}F(x,y;u,v)&-\f12F(x,v;u,y)\\
-\f12F(x,y;u,v)&(L_{xv}+L_{uy})C_F+\f1{2N_c}F(x,v;u,y)\end{array}}
\ee with \be F(x,y;u,v)=L_{xu}-L_{xv}+L_{yv}-L_{yu}=g_s^4\int
d^2z[G(x-z)-G(y-z)][G(u-z)-G(v-z)]\ . \ee

The problem has been reduced to finding the eigenvalue $\lambda_\pm$ and eigenvectors
of $M.$ Indeed one has \be
\lr{\begin{array}{cc}a_n\\b_n\end{array}}=\lr{\prod_{i=1}^n\mu^{2}(z_i^+)}M^n
\lr{\begin{array}{cc}1\\0\end{array}}=\lr{\prod_{i=1}^n\mu^{2}(z_i^+)}
\lr{\begin{array}{cc}a_+\lambda_+^n+a_-\lambda_-^n\\b_+\lambda_+^n+b_-\lambda_-^n\end{array}}
\ee with $a_\pm$ and $b_\pm$ obtained from the eigenvectors of $M$.
The results read: \be
\lambda_\pm=\lr{\f{N_c}4-\f1{2N_c}}(L_{xy}+L_{uv})+\f{N_c}4(L_{xv}+L_{uy})+\f1{2N_c}F(x,y;u,v)
\pm\f{N_c}4\sqrt{\Delta} \ee \be a_\pm=\f{\sqrt{\Delta}\pm
F(x,u;y,v)}{2\sqrt{\Delta}}\ ,\hspace{0.5cm}
b_\pm=\mp\f{F(x,y;u,v)}{N_c\sqrt{\Delta}} \ee \be
\Delta=F^2(x,u;y,v)+\f4{N_c^2}F(x,y;u,v)F(x,v;u,y)\ . \ee Finally,
resuming the contributions yields \be \bra S_{q\bar q}(x,y)S_{q\bar
q}(u,v)\ket=\f{T}{N_c^2}
\left[N_c^2\lr{a_+e^{\mu^2\lambda_+}+a_-e^{\mu^2\lambda_-}}
+N_c\lr{b_+e^{\mu^2\lambda_+}+b_-e^{\mu^2\lambda_-}}\right] \ee with
\be \mu^2=\int dz^+\mu^2(z^+)\ . \ee

One can write the final result in the following form \bea \bra
S_{q\bar q}(x,y)S_{q\bar q}(u,v)\ket=
\underbrace{e^{-\f{C_F}2[F(x-y)+F(u-v)]}}_{=\bra S_{q\bar
q}(x,y)\ket\bra S_{q\bar q}(u,v)\ket}\
\left[\lr{\f{F(x,u;y,v)\!+\!\sqrt{\Delta}}{2\sqrt{\Delta}}
-\f{F(x,y;u,v)}{N_c^2\sqrt{\Delta}}}e^{\f{N_c}4\mu^2\sqrt{\Delta}}\right.\nn\\\left.-
\lr{\f{F(x,u;y,v)\!-\!\sqrt{\Delta}}{2\sqrt{\Delta}}
-\f{F(x,y;u,v)}{N_c^2\sqrt{\Delta}}}e^{-\f{N_c}4\mu^2\sqrt{\Delta}}\right]
\ e^{-\f{N_c}4\mu^2F(x,u;y,v)+\f1{2N_c}\mu^2F(x,y;u,v)}\ . \label{mainresult}\eea It
is given in terms of a single function \be
F(x-y)=\mu^2(L_{xx}+L_{yy}-2L_{xy})=g_s^4\mu^2\int
d^2z\left[G(x-z)-G(y-z)\right]^2\ . \ee Indeed also one has \be
-2\mu^2F(x,y;u,v)=F(x-u)+F(y-v)-F(x-v)-F(y-u)\ . \ee Note that for
$x=y$ or $u=v,$ $F(x,y;u,v)=0$ and one recovers the single dipole
average.

In the function $F(r)$, the infrared cutoff $\Lambda_{QCD}$ only
enters through a logarithm as expected. In the
$|r|\Lambda_{QCD}\!\ll\!1$ limit, one has \bea
\f{C_F}2F(r)&=&\f{g_S^4C_F}{2\pi}\lr{\int dz^+\ \mu^2(z^+)}
\int_{\Lambda_{QCD}}^\infty dk\ \f{1-J_0(k|r|)}{k^3}\nn\\
&\simeq& \f{r^2}4\ \underbrace{\f{g_S^4C_F}{4\pi}\lr{\int
dz^+\mu^2(z^+)} \log\lr{\f1{r^2\Lambda_{QCD}^2}}}_{\equiv Q_s^2(r)}\
.\label{qsatmv}\eea This is the standard definition of the
saturation scale in the MV model. In the following, we will neglect
the logarithmic dependence of $Q_s,$ meaning \be
F(r)=\f{Q_s^2}{2C_F}r^2,\hspace{0.5cm}
\mu^2F(x,y;u,v)=\f{Q_s^2}{2C_F}(x-y)\cdot(u-v). \ee Finally, note
that this model is easily generalized to hot matter. It is showed in
Ref.\cite{bin&fabio} that by only keeping terms up to first order in
$k_\perp/k_z$, the gluon propagator in light-cone gauge for a color
charge propagating along the light cone in hot matter is
\begin{equation}
\begin{split}
D^{\mu\nu}_R\simeq-\frac{i}{k_\rho k^\rho-\mu_D^2}\left[g^{\mu\nu}-\frac{\eta^\mu
k^\nu+\eta^\nu k^\mu}{\eta_\mu k^\mu}\right],
\end{split}
\end{equation}
where $\mu_D^2=\f1{6}g^2T^2(N_c+\f{N_f}{2})$ and
$\eta^\mu=\f1{\sqrt{2}}(1,0,0,-1)$. To generalize the MV model to
hot matter, we treat the valence parton distribution as recoilless
color sources which are localized along the light cone, and in the
eikonal approximation, the only change of this model is to redefine
$G(x)$ as \be G(x)=\!\int\!\f{d^2k}{{(2\pi)^2}} \f{e^{ik\cdot
x}}{k^2+\mu_D^2}\ ,\ee and the infrared cutoff $\Lambda_{QCD}$ is
replaced by the Debye mass of the thermal plasma. In the next
section, we will see our generalized MV model is equivalent to the
GW model in light-cone coordinates in light-cone gauge.

\subsection{The quarkonium case}

Let us consider the situation where the two dipoles have identical
center of mass: \be
x=X+\f{r}2\hspace{0.5cm}y=X-\f{r}2\hspace{0.5cm}u=X-\f{r'}2\hspace{0.5cm}v=X+\f{r'}2\
. \ee Then one finds \be F(x,y;u,v)=-\f{Q_s^2}{2C_F}r\cdot
r'\hspace{0.5cm}F(x,u;y,v)=-\f{Q_s^2}{8C_F}(r+r')^2 \ee which yields
\bea \bra S_{q\bar q}(r)S_{q\bar
q}(r')\ket=e^{-\f{Q_s^2}4(r^2+r'^2)}\
\left[\lr{\f{-(r+r')^2/4\!+\!\sqrt{\Delta'}}{2\sqrt{\Delta'}}
+\f{r\cdot r'}{N_c^2\sqrt{\Delta'}}}e^{\f{Q_s^2}4
\f{\sqrt{\Delta'}}{1\!-\!1/N_c^2}}\right.\nn\\\left.
+\lr{\f{(r+r')^2/4\!+\!\sqrt{\Delta'}}{2\sqrt{\Delta'}} -\f{r\cdot
r'}{N_c^2\sqrt{\Delta'}}}e^{-\f{Q_s^2}4\f{\sqrt{\Delta'}}{1\!-\!1/N_c^2}}\right]\
e^{\f{Q_s^2}{16}\f{(r+r')^2}{1\!-\!1/N_c^2}-\f{Q_s^2}2\f{r\cdot
r'}{N_c^2\!-\!1}} \label{quarkonium}\eea with \be \Delta'=\f{(r+r')^4}{16}-\f{r\cdot
r'}{N_c^2}(r-r')^2\ .\ee

The large$-N_c$ limit gives:
\bea \bra S_{q\bar q}(r)S_{q\bar
q}(r')\ket=e^{-\f{Q_s^2}4(r^2+r'^2)}\
\left[1-\f{16}{N_c^2}\f{(r\cdot
r')^2}{(r+r')^4}\lr{1+\f{Q_s^2(r+r')^2}8}\right]
\nn\\+\f{16}{N_c^2}\f{(r\cdot
r')^2}{(r+r')^4}e^{-\f{Q_s^2}{8}(r-r')^2}\ . \eea

\subsection{The heavy meson case}

We now consider the situation where both antiquarks are at the same
position: \be
x=X+r\hspace{0.5cm}y=X\hspace{0.5cm}u=X\hspace{0.5cm}v=X+r'\ . \ee
Then one finds \be F(x,y;u,v)=F(x,u;y,v)=-\f{Q_s^2}{2C_F}r\cdot r'
\ee which significantly simplifies the result: \be \bra S_{q\bar
q}(r)S_{q\bar q}(r')\ket=e^{-\f{Q_s^2}4(r^2+r'^2)}
\left[\f1{N_c^2}e^{\f{Q_s^2}2r\cdot r'}
+\lr{1-\f1{N_c^2}}e^{-\f{Q_s^2}2\f{r\cdot r'}{N_c^2\!-\!1}} \right]\
. \ee

The large $N_c-$limit is simply

\bea \bra S_{q\bar q}(r)S_{q\bar
q}(r')\ket&=&e^{-\f{Q_s^2}4(r^2+r'^2)}
\left[1+\f1{N_c^2}\lr{e^{\f{Q_s^2}2r\cdot r'} -1-\f{Q_s^2}2r\cdot
r'} \right]\\&=&e^{-\f{Q_s^2}4(r^2+r'^2)}
-\f1{N_c^2}e^{-\f{Q_s^2}4(r^2+r'^2)}\lr{1+\f{Q_s^2}2r\cdot r'}
+\f1{N_c^2}e^{-\f{Q_s^2}4(r-r')^2} \eea

\section{Multiple scattering of a color singlet dipole in the Gyulassy-Wang model} \label{GW}

In this section, we use the GW model \cite{gw} to deal with the
multiple scatterings of a color singlet dipole in hot QCD matter.
The special case of heavy-meson dissociation due to multiple
scattering is addressed in \cite{Adil:2006ra}, however the color
structure is ignored in that analysis. In the following, we will
show that the GW model gives the same results as MV model in terms
of the saturation momentum $Q_s$ in the eikonal approximation.

\subsection{Introduction to the GW model}

In the GW model, the medium is modeled by an interaction Hamiltonian with $N\rightarrow\infty$ scatterers:
\begin{equation}
H_I(t)=\sum\limits_{i=1}^N\sum\limits_{a_i=1}^{N_c^2-1}\int
d^3x\left[
\Psi^\dagger_q(x)T^{a_i}V_i(x)\Psi_q(x)+\Psi^\dagger_{\bar{q}}(x)T^{a_i}V_i(x)\Psi_{\bar{q}}(x)\right]\equiv
H_q(t)+H_{\bar{q}}(t)\ .
\end{equation}
The screened potential
\be
V_i(\vec{x})=\frac{-\alpha}{|\vec{x}-\vec{z}_i|}e^{-\mu_D|\vec{x}-\vec{z}_i|}\ ,
\ee
or $V_i(\vec{q})=\frac{-4\pi\alpha}{q^2+\mu_D^2}e^{-i\vec{q}\cdot\vec{z}_i}$
in momentum space, is characterized by the Debye mass $\mu_D,$ and describes the medium in the
situation $\mu_D\lambda\gg 1,$ with $\lambda$ the mean free path of a single (anti)quark. Even though we are dealing with multiple scattering of a color singlet $q\bar{q}$ dipole in this
paper, we still use the mean free path of a single (anti)quark and treat the
quark and antiquark in the dipole wave function as two individual
free particles. This approximation is good when the relevant length scale of the
medium $L\lesssim\gamma\tau$, the typical time scale in a meson \cite{bin&fabio}.
The assumption $1/\mu_D\ll\lambda$ means that the scatterers are independent of
each other and, therefore, completely uncorrelated.

\begin{figure}
\begin{center}
\includegraphics[width=15cm]{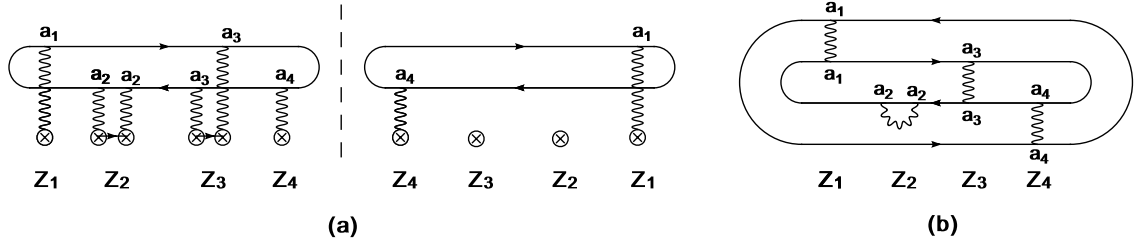}
\end{center}
\caption{The equivalence of the color factors between (a) and (b):
after the medium average, each pair of color indices $a_i$ of the
generators $T^{a_i}$ at the position $\vec{z}_i$ in the amplitude
and/or the conjugate amplitude is identified. Therefore, it has the
same color factor as the Feynman diagram (b). In the large $N_c$
limit, the amplitude of (b) is proportional to $C_F^2$ in contrast
with that in Fig.~\ref{ColorDiagramSS01}(b).}\label{ColorDiagram}
\end{figure}

The ensemble average over the transverse positions of the scatterers is defined as
\begin{equation}
\left<\cdot\cdot\cdot\right>\equiv\sum\limits_{i=1}^{N}\int\f{d^2z_{i
\perp}}A,
\end{equation}
and after the medium average, each pair of color indices $a_i$ of
the generators $T^{a_i}$ at the position $\vec{z}_i$ in the
amplitude and/or the conjugate amplitude is identified as
illustrated in Fig.~\ref{ColorDiagram}(a), which is equivalent to
(\ref{EnseAver}) in the MV model. 
This equivalence enables us to calculate the color structure in the same
algebraic way than showed in Sec. III. Instead, in this section we choose to calculate
the dipole-dipole correlator in the large-$N_c$ limit, which allows a diagrammatic analysis.
We shall calculate the probability $P=\sum\limits_{n=0}^\infty P^{(n)}$ in the eikonal
approximation with
\begin{equation}
P^{(n)}=\f1A\int\f{d^2P^\prime_\perp}{(2\pi)^2}\left<|M^{(n)}|^2\right>\ ,
\label{Pn}
\end{equation}
the contribution to $P$ from those diagrams with $n$ scattering centers
scattering with the dipole in the amplitude and/or the conjugate
amplitude.

Among $N$ scatterers, we have 
$C_N^n=\f{N!}{n!(N-n)!}\simeq\f{N^n}{n!}$ choices of the $n$
scatterers which scatter with the dipole in the amplitude and/or the
conjugate amplitude. In the eikonal approximation, each choice of
those $n$ scatterers will give the same contribution to $P^{(n)}$.
Therefore, let us assume that the dipole, moving along the $+$
direction, is scattered by the first $n$ scatterers which are located at
$\vec{z}_i,i=1,\cdot\cdot\cdot,n$ respectively with
$z^+_i>z^+_{i-1}$. In the following, we enumerate the $n$ scatterers
by $z_i$ and by $z_i>z_j$ we mean $z^+_i>z^+_j$.
Focusing on the color factor of each term contributing to $P^{(n)}$,
there exists a Feynman diagram with the same color factor. An
example is showed in Fig.~\ref{ColorDiagram}. We will use this
correspondence to evaluate $P$ by a diagrammatic analysis at large
$N_c$. Up to $\mathcal{O}(\alpha^2)$, we will evaluate separately
the following two cases:
\begin{itemize}
\item single-scattering diagrams such as those shown in Fig.~\ref{ColorDiagramSS01}(a).
In this case, since the $n$ scattering centers all scatter with the
dipole by single scattering in the amplitude, they must also scatter
with it in the conjugate amplitude. This corresponds to processes in which 
the target scatters inelastically.
\item double-scattering diagrams such as those shown in Fig.~\ref{ColorDiagram}(a).
In this case, if the scatterer $z_i$ undergoes double scattering in the amplitude,
there must be no scattering between it and the dipole in the conjugate amplitude
and vice versa. This corresponds to processes in which the target scatters elastically.
\end{itemize}

\subsection{The evaluation of single-scattering diagrams}

\begin{figure}
\begin{center}
\includegraphics[width=15cm]{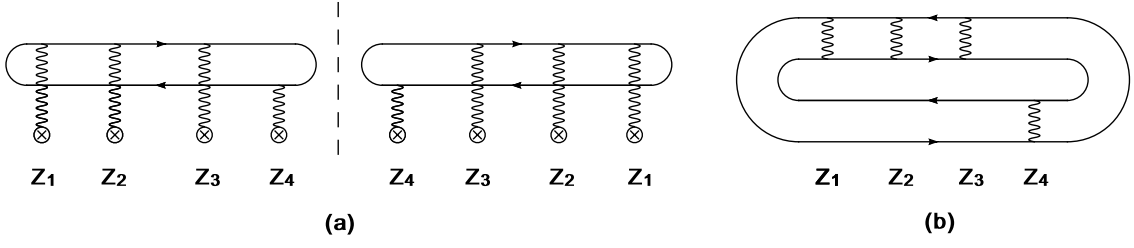}
\end{center}
\caption{An example of single-scattering diagram of the leading
order in $N_c$: The contribution of (a) to $P^{(n)}$ has the same
color factor as the amplitude of (b), which is proportional to
$C_F^4$ in the large $N_c$ limit.} \label{ColorDiagramSS01}
\end{figure}

The amplitude for the dipole to undergo $n$ single scatterings with
those $n$ scatterers $M^{(n)}_{s}$ is a sum of $2^n$ corresponding
diagrams since the gluon line from each scatterer can hook either on
the quark or the antiquark line. If the gluon line from the
scatterer $z_i$ hooks on the quark or the antiquark line, the
corresponding amplitude picks up a phase $e^{i\f{m_{\bar{q}}}M
q_{\perp i}\cdot x_{\perp}}=e^{iq_{\perp
i}\cdot(x_{q\perp}-X_\perp)}$ or $e^{-i\f{m_{q}}M q_{\perp i}\cdot
x_{\perp}}=e^{iq_{\perp i}\cdot(x_{\bar{q}\perp}-X_\perp)}$
respectively, where $q_{\perp i}$ denotes the transverse momentum of the gluon line.
In the mean time, the color matrix $T^{z_i}$ is put in
the quark part or antiquark part of the trace of the $n$ color
matrices accordingly. For example, if the scatterers
$z_{a_1}<\cdot\cdot\cdot<z_{a_m}$ and
$z_{b_1}<\cdot\cdot\cdot<z_{b_{n-m}}$ scatter with the quark and
antiquark respectively, we have $\mbox{Tr}\left[T^{z_{a_m}}\cdot\cdot\cdot
T^{z_{a_1}}T^{z_{b_1}}\cdot\cdot\cdot T^{z_{b_{n-m}}}\right]$
in the amplitude of the corresponding diagram. In general, we have
\begin{equation}
\begin{split}
M^{(n)}_{s}=&{ (i)^{n} }\f1{N_c}\int d^2x_{\perp}
\Phi^\lambda_{fi}(x_\perp)\int\prod\limits_{i=1}^{n-1}\left[
\f{d^2q_{\perp i}}{(2\pi)^2}\right]\\
\times&\prod\limits_{i=1}^{n}\left[\f{ 4\pi\alpha e^{-iq_{\perp
i}\cdot z_{\perp i}}}{q_{\perp i}^2+\mu_D^2}\left(e^{i\f{m_{\bar{q}}}M
q_{\perp i}\cdot
x_{\perp}}T^{a_i}_{c_{i+1}c_i}\delta_{d_{i}d_{i+1}}-e^{-i\f{m_{q}}M
q_{\perp i}\cdot
x_{\perp}}T^{a_i}_{d_{i}d_{i+1}}\delta_{c_{i}c_{i+1}}\right)\right],
\end{split}
\end{equation}
with  $c_1=d_1$, $d_{n+1}=c_{n+1}$ and $M=m_{q}+m_{\bar{q}}$.

In fact, it is easy to show that we only have $2^{n-2}$ different
traces of color matrices in $M^{(n)}_s$ which correspond
respectively to the $2^{n-2}$ different hookings of the gluon lines
from the $n-2$ scatterers in between $z_1$ and $z_n$ on the quark
and antiquark lines. Generally, we write the trace in the following
form
\begin{equation}
\mbox{Tr}^{(n,m)}\equiv
\mbox{Tr}\left[T^{z_n}T^{z_{a_m}}\cdot\cdot\cdot
T^{z_{a_1}}T^{z_1}T^{z_{b_1}}\cdot\cdot\cdot
T^{z_{b_{n-m-2}}}\right],
\end{equation}
and the corresponding four diagrams give to $M^{(n)}_s$ a
contribution
\begin{equation}
\begin{split}
M^{(n,m)}_{s}=&{ (i)^{n} }\f{(-1)^{n-m-2}}{N_c}\mbox{Tr}^{(n,m)}\int
d^2x_{\perp}
\Phi^\lambda_{fi}(x_\perp)\int\prod\limits_{i=1}^{n-1}\left[
\f{d^2q_{\perp i}}{(2\pi)^2}\right]\\
\times&\prod\limits_{i=1}^{n}\left[\f{4\pi\alpha e^{-iq_{\perp
i}\cdot z_{\perp i}}}{q_{\perp
i}^2+\mu_D^2}\right]e^{i\f{m_{\bar{q}}}M \sum\limits_{j=1}^m q_{\perp
a_j}\cdot x_{\perp}-i\f{m_{q}}M
\sum\limits_{j=1}^{n-m-2} q_{\perp b_j}\cdot x_{\perp}}\\
\times&\left(e^{i\f{m_{\bar{q}}}M q_{\perp 1}\cdot
x_{\perp}}-e^{-i\f{m_{q}}M q_{\perp 1}\cdot
x_{\perp}}\right)\left(e^{i\f{m_{\bar{q}}}M q_{\perp n}\cdot
x_{\perp}}-e^{-i\f{m_{q}}M q_{\perp n}\cdot
x_{\perp}}\right).\label{MnDleading}
\end{split}
\end{equation}

\begin{figure}
\begin{center}
\includegraphics[width=15cm]{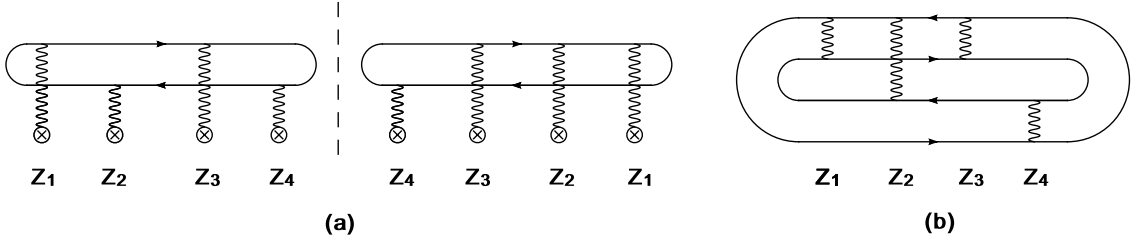}
\end{center}
\caption{An example for $1/N_c^2$ suppressed diagram of
single-scatterings to $P^{(n)}$: in (a) the gluon line from the
scatterer $z_2$ hooks on different fermion lines in the amplitude
and conjugate amplitude. It has the same color factor as the Feynman
diagram (b), which is non-planar and $1/N_c^2$
suppressed relative to Fig.~\ref{ColorDiagramSS01}(b) \cite{Witten:1979kh}.}\label{ColorDiagramSS}
\end{figure}

In the large $N_c$ limit, to evaluate
\begin{equation}
P_s^{(n)}=C^n_N\f1A\int\f{d^2P^\prime_\perp}{(2\pi)^2}\left<|M_s^{(n)}|^2\right>,\label{Pdn}
\end{equation}
we only need to take into account those diagrams with each pair of
the $n-2$ scatterers in between $z_1$ and $z_n$ hooking on the same
fermion line in the amplitude and conjugate amplitude. For example,
in the case $n=4$, the contribution to $P_D^{(n)}$ from the diagram
Fig.~\ref{ColorDiagramSS01}(a) has the same color factor as Fig.~\ref{ColorDiagramSS01}(b) $\propto C_F^4$, while the color factor
of the contribution from Fig.~\ref{ColorDiagramSS}(a) correspond to
a non-planar diagram Fig.~\ref{ColorDiagramSS}(b) and, therefore,
are $1/N_c^2$ suppressed relative to Fig.~\ref{ColorDiagramSS01}(b) \cite{Witten:1979kh}. 
This means that we only need to count in the $2^{n-2}$ products of $M^{(n,m)}_{D}$ and
their conjugates, which have a unique color factor
\begin{equation}
\mbox{Tr}\left[T^{z_n}\cdot\cdot\cdot
T^{z_1}\right]\mbox{Tr}\left[T^{z_1}\cdot\cdot\cdot
T^{z_n}\right]=C_F^n+(-1)^n\f{N_c^2-1}{2^nN_c^n}\simeq C_F^n.
\end{equation}
Therefore, for arbitrary $n\geq2$, we have in the large $N_c$ limit
\begin{equation}
\begin{split}
P_s^{(n)}=&\f1{n!}P_2(Z+2\chi)^{n-2},
\end{split}
\end{equation}
where
\begin{equation}
\begin{split}
P_2&\equiv\f1{N_c^2}(L\rho\sigma)^2\int d^2x_{\perp}
d^2x_{\perp}^\prime\Phi^\lambda_{fi}(x_\perp)\Phi^{\lambda*}_{fi}(x^\prime_\perp)\\
&\times\left[\int d^2q_{\perp}\f{\mu_D^2}{\pi(q_{\perp}^2+\mu_D^2)^2}
\left(e^{i\f{m_{\bar{q}}}M q_{\perp}\cdot x_{\perp}}
-e^{-i\f{m_{q}}M q_{\perp}\cdot x_{\perp}}\right)
\left(e^{-i\f{m_{\bar{q}}}M q_{\perp}\cdot x^\prime_{\perp}}
-e^{i\f{m_{q}}M q_{\perp}\cdot x^\prime_{\perp}}\right)\right]^2\\
&\simeq\f1{N_c^2}\int d^2x_{\perp} d^2x_{\perp}^\prime
\Phi^\lambda_{fi}(x_\perp)\Phi^{\lambda*}_{fi}(x^\prime_\perp)
\f14 Q_s^4(x_\perp\cdot x_\perp^\prime)^2\ ,
\end{split}\label{p2}
\end{equation}
and
\begin{equation}
\begin{split}
Z&\equiv\int
d^2q_{\perp}\f{L\rho\sigma\mu_D^2}{\pi(q_{\perp}^2+\mu_D^2)^2}
\left(e^{i\f{m_{\bar{q}}}M q_{\perp}\cdot
(x_{\perp}-x_{\perp}^\prime)} + e^{-i\f{m_{q}}M
q_{\perp}\cdot(x_{\perp}-x_{\perp}^\prime)}-2\right)\ .
\end{split}
\end{equation}

We have introduced the number density in the transverse plane
$\f NA=L\rho$ with $L$ the plasma length and $\rho$ the number density of scatterers, the cross section of the (anti)quark undergoing a single scattering $\sigma=\f{4\pi\alpha^2C_F}{\mu_D^2}$, and the average scattering number $\chi=\f{L}{\lambda}=L\rho\sigma.$
Finally,
\be Q_s^2=L\rho\sigma\int\limits_{|q_\perp|<1/|x_\perp|}
d^2q_{\perp}\f{\mu_D^2}{\pi(q_{\perp}^2+\mu_D^2)^2}q_{\perp}^2\label{QsGW}
\ee is the plasma saturation momentum squared, which is a characteristic property of the QCD medium. It is different from
the traditional saturation momentum introduced in the previous section (which characterizes the small$-x$ part of a hadronic
wave function), but when writting \eqref{QsGW} in terms of the gluon density per unit of transverse area in the plasma \cite{Baier:1996kr}, 
one finds the same expression than when writting the saturation scale in terms of the gluon density per unit of transverse area in a hadron. 
And indeed, one can check that formula \eqref{qsatmv} is recovered in the $|x_\perp|\mu_D\ll1$ limit if one replaces $\mu_D$ with $\Lambda_{QCD}$ in (\ref{QsGW}). In terms of the traditional jet quenching parameter $\hat{q},$ one has $Q_s^2=hat{q}L.$
Note also that in formula \eqref{p2}, we have assumed that $|q_\perp\cdot x_\perp|\ll 1.$ This is justified because the typical values are $|q_\perp|\sim T$ while $|x_\perp|\sim a_B,$ the meson size. We always assume a situation where $Ta_B\ll1,$ where multiple scatterings are important.

\begin{figure}
\begin{center}
\includegraphics[width=10cm]{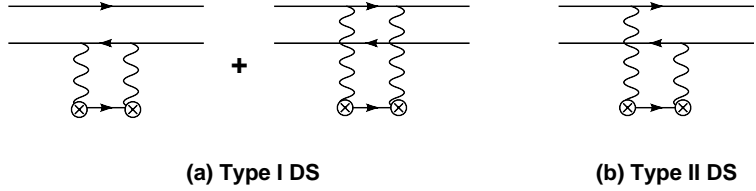}
\end{center}
\caption{Two types of double-scatterings for each scatterer in the
amplitude: after the medium average the color factor of (a) is
trivially $C_F$ while the color factor of (b) depends on its
relative position with respect to those for single-scatterings. If
it is not in between any two of single-scatterings, it also has a
color factor $C_F$.}\label{FourTypeScat}
\end{figure}

\subsection{The evaluation of double-scattering diagrams}

In all the diagrams, we only need to evaluate two types of
double-scatterings as showed in Fig.~\ref{FourTypeScat}. Each type I
double-scatterings and its conjugate give a contribution $-2\chi$ to
$P^{(n)}$. This is easily read out from the corresponding diagrams
such as the type I double-scattering at $z_2$ in
Fig.~\ref{ColorDiagram}(a). Those diagrams with at least one type II
double-scattering in between two single-scatterings are $1/N_c^2$
suppressed. For example, the contribution to $P^{(n)}$ corresponding
to Fig.~\ref{ColorDiagram}(a) $\propto C_F^2$ is $1/N_c^2$
suppressed compared with that in Fig.~\ref{ColorDiagramSS01}(a) in
the large $N_c$ limit. Otherwise, each type II double-scattering and
its conjugate give to $P^{(n)}$ a contribution
\begin{equation}
\begin{split}
Y+2\chi&\equiv\int
d^2q_{\perp}\f{L\rho\sigma\mu_D^2}{\pi(q_{\perp}^2+\mu_D^2)^2}
\left(e^{iq_{\perp}\cdot x_{\perp}}+e^{-iq_{\perp}\cdot
x_{\perp}^\prime}\right),
\end{split}
\end{equation}
with
\begin{equation}
\begin{split}
Y\simeq-\f14 Q_s^2(x_\perp^2+x_\perp^{\prime2}).
\end{split}
\end{equation}
If the dipole interacts with the medium only through
double-scatterings, that is, the target is not excited, we get the
probability for the so-called elastic process
\begin{equation}
\begin{split}
P^\lambda_{el}&=\int d^2x_{\perp} d^2x_{\perp}^\prime
\Phi^\lambda_{fi}(x_\perp)\Phi^{\lambda
*}_{fi}(x^\prime_\perp)\sum\limits_{m_I,m_{II}=0}^\infty
\f1{m_I!m_{II}!}(-2\chi)^{m_I}(Y+2\chi)^{m_{II}}\\
&=\int d^2x_{\perp} d^2x_{\perp}^\prime
\Phi^\lambda_{fi}(x_\perp)\Phi^{\lambda
*}_{fi}(x^\prime_\perp)e^{-\f14 Q_s^2(x_\perp^2+x_\perp^{\prime2})}.
\end{split}\label{elastic}
\end{equation}

\subsection{The survival probability from inelastic multiple scatterings}

Now, we are ready to evaluate $P_{inel}^{(n)}$ up to order
$1/N_c^2$, the probability for the transition of an initial state
$\varphi_{i}$ to a final sate $\varphi_{f}$ undergoing at least
$n\geq2$ single-scatterings with the medium excited. Let us assume
that there are $m_I$ type I double-scatterings and $m_{II}$ type II
double-scatterings among total $n$ scatterings. First, we have
$C_N^n=\f{N!}{n!(N-n)!}\simeq\f{N^n}{n!}$ choices of the $n$
scatterers. Then out of those $n$ scatterers, we have $C_{n}^{n-m_I}$ possible
choices of $(n-m_I)$ scatterers undergoing $(n-m_I-m_{II})$
single-scatterings and $m_{II}$ type II double-scatterings. Since we only
need to count in the diagrams with no type II double-scattering in
between any two of all the $(n-m_I-m_{II})$ single-scatterings, we only have
$(m_{II}+1)$ choices of the $m_{II}$ scatterers undergoing type II double-scattering.

Therefore, for each group $(n,m_I,m_{II})$, we have
\begin{equation}
\begin{split}
P^{(n,m_I,m_{II})}&=\f1{n!}
C_n^{n-m_I}(m_{II}+1)P_2(-2\chi)^{m_I}(Y+2\chi)^{m_{II}}(Z+2\chi)^{n-m_I-m_{II}-2}\\
&=\f{m_{II}+1}{m_I!(n-m_I)!}P_2(-2\chi)^{m_I}(Y+2\chi)^{m_{II}}(Z+2\chi)^{n-m_I-m_{II}-2}\ ,
\end{split}
\end{equation}
which leads to
\begin{equation}
P_{inel}^{(n)}=\sum\limits_{m_{II}=0}^{n-2}\sum\limits_{m_{I}=0}^{n-2-m_{II}}P^{(n,m_I,m_{II})}
=\f1{n!}\sum\limits_{m=0}^{n-2}(m+1)Y^mZ^{n-m-2}\ ,
\end{equation}
and finally
\begin{equation}
P_{inel}=\sum\limits_{n=2}^\infty
P_{inel}^{(n)}=\f{P_2}{(Y-Z)^2}\left[e^Z+(Y-Z-1)e^Y\right]\ .
\end{equation}

We can now discuss the two situations of Fig.~\ref{TwoCaseofMC} (we recall that
$x_\perp=x_{q\perp}-x_{\bar{q}\perp}$ and 
$x^\prime_\perp=x^\prime_{q\perp}-x^\prime_{\bar{q}\perp}$).
\begin{itemize}
\item (a) In the quarkonium case: $m_{\bar{q}}\simeq m_{q}$ and
$X_\perp\simeq(x_{q\perp}+x_{\bar{q}\perp})/2=(x^\prime_{q\perp}+x^\prime_{\bar{q}\perp})/2,$
this leads to $Z=-\f{Q_s^2}8(x_\perp-x_\perp^\prime)^2$ and
\begin{equation}
\begin{split}
P^\lambda_{inel}&=\f1{N_c^2}\int d^2x_{\perp} d^2x_{\perp}^\prime
\Phi^\lambda_{fi}(x_\perp)\Phi^{\lambda*}_{fi}(x^\prime_\perp)\\
&\times\left\{\f{16(x_\perp\cdot
x_\perp^\prime)^2}{(x_\perp+x_\perp^{\prime})^4} e^{-\f18
Q_s^2(x_\perp-x_\perp^\prime)^2}- \f{2(x_\perp\cdot
x_\perp^\prime)^2}{(x_\perp+x_\perp^{\prime})^2}
\left[Q^2_s+\f8{(x_\perp+x_\perp^{\prime})^2}\right]e^{-\f14
Q_s^2(x_\perp^2+x_\perp^{\prime2})}\right\}.
\end{split}\label{inelquarkonium}
\end{equation}
\item (b) In the heavy-meson case: $m_{\bar{q}}\gg m_{q},$ and
$X_\perp\simeq x_{\bar{q}\perp}=x^\prime_{\bar{q}\perp},$ this leads to
$Z=-\f{Q^2_s}4(x_\perp-x_\perp^\prime)^2$ and
\begin{equation}
\begin{split}
P^\lambda_{inel}&=\f1{N_c^2}\int d^2x_{\perp} d^2x_{\perp}^\prime
\Phi^\lambda_{fi}(x_\perp)\Phi^{\lambda*}_{fi}(x^\prime_\perp)\\&\times
\left[e^{-\f14 Q_s^2(x_\perp-x_\perp^\prime)^2}\!-\!\left(1+\f12
Q^2_sx_\perp\cdot x_\perp^\prime\right)e^{-\f14
Q_s^2(x_\perp^2+x_\perp^{\prime2})}\right].
\label{inelhm}
\end{split}
\end{equation}
\end{itemize}

In both cases, we obtain the same survival probability
\be
P^\lambda_{el}+P^\lambda_{inel}=\int d^2x_{\perp} d^2x_{\perp}^\prime
\Phi^\lambda_{fi}(x_\perp)\Phi^{\lambda*}_{fi}(x^\prime_\perp)
\bra S_{q\bar q}(x_\perp)S_{q\bar q}(x^\prime_\perp)\ket
\ee
as in the previous section, the differences enter through the saturation scales only. The infrared scale is $\Lambda_{QCD}$ in the cold QCD matter case, and $\mu_D$ for hot QCD matter.

\section{Phenomenological implications}

It is very interesting to see how the two different approaches, for cold and hot matter, yield the same result when expressed in terms of the appropriate saturation scale. This fact allows us to make general comments about bound state dissociation and vector meson production leaving all the medium dependence in only one parameter. From the calculations shown in previous sections we can easily see
that in the large $N_{c}$ limit, the main contribution to the dipole-dipole correlator comes from elastic processes. The contribution coming from inelastic processes is $1/N_c^2$ suppressed, and is therefore 10 \% smaller in general. However, the different dependences of the subleading contributions on the size and orientation of the dipoles can play an important role in determining when these contributions become comparable to the one from the elastic case. For instance, we notice that the elastic correlator (as shown in \eqref{elastic}) is insensitive to the relative orientation of the dipoles but the inelastic contribution is not.

First, it is worth noticing that the inelastic contribution goes to zero as $Q_{s}^4$ when $Q_{s}$ goes to zero both for the heavy meson and the quarkonium case. In this limit which corresponds to turning off the interaction, elastic dissociation goes to zero only as $Q_s^2,$ therefore in order to get a large contribution from inelastic processes $Q_{s}$ should be large enough. This observation has to be taken with caution since $Q_{s}$ is present in all of the exponentials. A too large $Q_{s}$ induces a big suppression in the dipole-dipole correlator (elastic or inelastic) except for very small dipole sizes, which don't give a big contribution when the wave functions are included (Eq. (\ref{Probability}), (\ref{cross-section})).

With the above considerations in mind, we can easily notice that the terms in the inelastic part of the correlator with the exponential
factor $e^{-\frac{Q_{s}^{2}}{4}(x_{\perp}^{2}+x_{\perp}^{\prime 2})}$ in Eqs. (\ref{inelquarkonium}) and (\ref{inelhm}) don't give a
large contribution, since they only become comparable with the elastic part in a region heavily suppressed by the exponential
factor and the wave functions. Following this observation we turn our attention to the terms with the exponential factors of the form
$e^{-\frac{Q_{s}^{2}}{4}(x_{\perp}-x'_{\perp})^{2}}$ and $e^{-\frac{Q_{s}^{2}}{8}(x_{\perp}-x'_{\perp})^{2}}$. Unlike the
other terms, these exponentials are not necessarily small for large dipoles and don't exhibit the property of color transparency, which is characteristic of the elastic terms. As long as the two dipoles are aligned and similar in size, the first term in both expressions for the inelastic part can overcome the $1/N_{c}^{2}$ suppression and become comparable with the elastic part.

More precise comparisons of the relative sizes for different dipole configurations require a more intricate numerical analysis which is left for future work, especially the result of the interplay between the size dependences of the meson wave functions and those of the dipole-dipole correlator should be investigated numerically. Still, let us make more specific comments concerning different physical situations where we expect our results to be relevant.

\subsection{Super-penetration of quarkonia}

\begin{figure}
\begin{center}
\includegraphics[width=10cm]{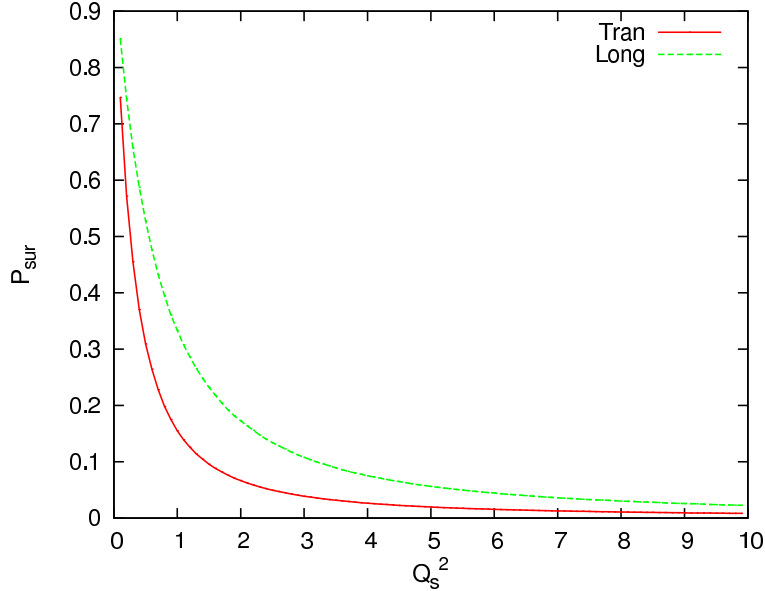}
\end{center}
\caption{Survival probabilities for $J/\psi$ mesons, using ``light-cone gaussian'' wave functions and the dipole-dipole correlator \eqref{quarkonium}.}
\label{probabvsQs}
\end{figure}

As an example of a full numerical calculation, we considered the $J/\psi$ survival probability. The light-cone wave functions of
Ref.~\cite{lc-jpsi} were used to obtain the probability plotted in Fig.~\ref{probabvsQs} as a function of $Q_{s}^{2}.$
Current measurements in heavy-ion collisions at RHIC suggest a value of $Q_{s}^{2} \sim 1 \text{ GeV}^{2}$ for a gold nucleus and $Q_{s}^{2}=\hat{q}L \sim 20 \text{ GeV}^{2}$ for a plasma with a size of a few fermis for central collisions. At the LHC, these values
can be increased by a factor of order 3, given the accessibility to partons with smaller momentum fraction in the nuclear wave function, or higher plasma temperatures in the hot matter case. We can anticipate that the results at the LHC should be qualitatively different since the survival probability decreases by about one order of magnitude.

We also observe on Fig.~\ref{probabvsQs} that the survival probability decreases as $1/Q_s^2$ for large values of $Q_s.$ In the hot matter case, this corresponds to a $1/L$ behavior in terms of the plasma length. This dependence, already derived previously \cite{HiroFujii} (and in \cite{hlm,FujiiMatsui} for an $SU(2)$ color group), is in contrast with the $e^{-L}$ decrease usually assumed. 
Such an exponential dependence is valid when the dipole undergoes successful independent scatterings in the medium, while the power-law we obtain is due to the coherence of the multiple scatterings, and both the elastic and inelastic contributions display this {\it super-penetration} feature \cite{superpen}. A similar scaling was also observed when considering the suppressed production of heavy quarks off cold nuclear matter \cite{fgv}.

\subsection{Collisional dissociation of heavy mesons}

While for light hadrons, partonic energy-loss calculations can describe the suppressed production of high$-p_T$
particles in nucleus-nucleus collisions at RHIC, the suppression is underestimated in the case of $D$ and $B$ mesons
(or rather single non-photonic electrons coming from the decays) \cite{collEloss}. The discrepancy is due to the small quenching of $B$-mesons, which dominate the high-$p_T$  single electron yields. However, the standard calculations assume that the hard parton hadronizes outside the medium, having fully traversed the region of dense nuclear matter, and lost energy via radiative and collisional processes. But for heavy mesons, due to the significantly smaller formation times, this assumption does not appear to be justified.
In fact, it was shown that including collisional dissociation goes in the right direction \cite{Adil:2006ra}: contrary to calculations that emphasize radiative and collisional heavy quark energy loss, collisional dissociation predicts that $D$ and $B$ mesons are suppressed in a similar way at transverse momenta as low as $p_T \sim 10\ \mbox{GeV}.$

The framework of Ref.~\cite{Adil:2006ra} takes into account the competition between the fragmentation of $c$ and $b$ quarks and the medium-induced dissociation of the $D$ and $B$ mesons to evaluate the quenching in nucleus-nucleus collisions. The meson survival probability we have computed in the previous sections is an important ingredient of their analysis: it determines the dissociation time of the meson in the plasma. However the expression used in \cite{Adil:2006ra} ignores the color structure of the theory and therefore it would be interesting to redo the analysis with our expression, but this is beyond the scope of this work. Note that the time dependence of the plasma formation and evolution is easily implemented by varying the parameter $Q_s^2.$

\subsection{Diffractive vector meson production}

An experimental situation where elastic and inelastic processes can be distinguished is the diffractive production of vector mesons in deep inelastic scattering: $\g^* A \to V Y,$ where $A$ stands for the target nucleus and $Y$ for the final state it has dissociated into. In this process, the $q\bar q$ pair that the virtual photon has fluctuated into scatters off the nucleus before recombining into a vector meson. While the scattering involves a color-singlet exchange, leaving a rapidity gap in the final state, the nucleus can still scatter elastically ($Y=A,$ this is called coherent diffraction) or inelastically ({\it i.e.} break up, called incoherent diffraction). Kinematically, a low invariant mass of the system $Y$ corresponds to a large rapidity gap between that system and the vector meson (this also implies that the longitudinal momentum of the meson is close to that of the photon, which justifies using the eikonal approximation for this process),
which makes it possible in principle to keep track of the state of the target, and separate coherent and incoherent diffraction.

The cross-section is peaked at minimum momentum transfer where the elastic scattering dominates, but as the transfer of momentum gets larger, the role of the inelastic contribution increases and eventually it becomes dominant (typically for momenta bigger that the inverse nucleus size). The momentum transfer in this process is essentially the transverse momentum of the vector meson in the final state $P_\perp^\prime$, and as a function of $|t|=P_\perp^{\prime 2},$ the elastic contribution decreases exponentially while the inelastic contribution decreases only as a power law. This important difference was not discussed in this paper where only $P_\perp^\prime-$integrated quantities are analyzed. It deserves detailed studies which are left for future work, such as for instance the numerical analysis of our results and a comparison with data from HERA on diffractive vector meson production, with or without proton breakup. The case of deep inelastic scattering off a large nucleus should also be studied, and in this case the MV model we considered provides a natural framework, and a good starting point to implement the high-energy QCD evolution. Inclusive and diffractive structure functions have been calculated \cite{CGCeA}, but vector-meson production has yet to be addressed. At an electron-ion collider, when the momentum transfer is small enough for the nucleus to stay intact, then it will escape too close to the beam to be detectable; therefore the whole diffractive program will rely on our understanding of incoherent diffraction.

\section{Conclusions}

The main technical result of the paper is the 4-point function \eqref{mainresult}, computed in the MV model for cold nuclear matter, but also valid in the GW model for hot matter. This dipole-dipole correlator allows to compute the survival probability of a meson propagating in the presence QCD matter \eqref{Probability}, and the cross-section for the diffractive production of vector mesons in deep inelastic scattering off nuclei \eqref{cross-section}. In the hot nuclear matter case, only the large $N_c$ results were explicitely derived, with a diagrammatic approach which gave a more physical picture of what are the processes giving the main contribution to the dissociation of bound states. It also allowed a clear distinction between elastic and inelastic processes with respect to the target.

The two results for cold and hot matter are given by the same expression when written in terms of the saturation momentum of either the nucleus or the plasma (it is more common in this case to write $Q_s^2=\hat{q}L$). The medium dependence enters only through this parameter. While this allowed us to make general statements, the interplay between the meson wave functions and the dipole-dipole correlator should be analyzed numerically. Even though experimentally it is not easy to appreciate the difference between elastic and inelastic processes in heavy ion collisions, we consider our result to be interesting from the theoretical point of view since we address a realistic situation where the medium can be excited by the interaction with the meson.

We obtained that the survival probability of quarkonia traversing a hot QCD medium exhibits the super-penetration feature: it decrease as $1/L$ for large medium length $L.$ We also except our result to be relevant when considering the production of high-$p_T$ $D$ and $B$ mesons in nucleus-nucleus collisions, as these can be produced within the medium and their in-medium dissociation will contribute to the strong suppression observed at RHIC. This will be better investigated at the LHC where it should be possible to distinguish signals from charm and bottom quarks making it possible to differentiate between the two cases in a regime where we can safely rely on the eikonal approximation. Finally, an actual comparison of our results for elastic and inelastic processes is more accessible with the diffractive production of vector mesons in deep inelastic scattering off nuclei; they correspond to coherent and incoherent diffraction respectively which can be derived from our formulae.

\section*{Acknowledgments}

We would like to thank Prof. A.H. Mueller for numerous discussions and
helpful comments. BW is also grateful to Prof. B.-Q. Ma for useful suggestions.
CM is supported by the European Commission under the FP6 program,
contract No. MOIF-CT-2006-039860. BW is supported by China Scholarship Council.


\end{document}